\documentclass[12pt,longbibliography]{article}
\usepackage{amsfonts,epsfig,latexsym,amscd,amsmath,theorem,mathrsfs,color}

\graphicspath{{./images/}}
\DeclareMathOperator{\grad}{grad}

\newcommand{\EE}{\mathscr{E}} 
\newcommand{\CC}{\mathscr{C}}

\newcommand{\R}{{\mathbb{R}}}

\newcommand{\C}{{\mathbb{C}}}
\newcommand{\I}{{\mathbb{I}}}

\newcommand{\beq}{\begin{equation}}
\newcommand{\eeq}{\end{equation}}
\newcommand{\bea}{\begin{eqnarray}}
\newcommand{\eea}{\end{eqnarray}}
\newcommand{\ben}{\begin{eqnarray*}}
\newcommand{\een}{\end{eqnarray*}}
\newcommand{\bem}{\begin{enumerate}}
\newcommand{\eem}{\end{enumerate}}

\newcommand{\ra}{\rightarrow}

\newcommand{\cd}{\partial}
\newcommand{\wt}{\widetilde}

\def \d{\mathrm{d}}

\newcommand{\ignore}[1]{}

\newcommand{\epsvec}{\mbox{\boldmath{$\eps$}}}

\newcommand{\nvec}{\mbox{\boldmath{$n$}}}
\newcommand{\pvec}{\mbox{\boldmath{$p$}}}
\newcommand{\pvecs}{\mbox{\boldmath{$\scriptstyle p$}}}
\newcommand{\mvec}{\mbox{\boldmath{$m$}}}
\newcommand{\kvec}{\mbox{\boldmath{$k$}}}
\newcommand{\xvec}{\mbox{\boldmath{$x$}}}
\newcommand{\svec}{\mbox{\boldmath{$s$}}}
\newcommand{\uvec}{{\mbox{\boldmath{$u$}}}}
\newcommand{\uvecs}{{\mbox{\boldmath{$\scriptstyle u$}}}}
\newcommand{\vvec}{\mbox{\boldmath{$v$}}}
\newcommand{\wvec}{\mbox{\boldmath{$w$}}}
\newcommand{\evec}{{\mbox{\boldmath{$e$}}}}
\newcommand{\evecs}{{\mbox{\boldmath{$\scriptstyle e$}}}}
\newcommand{\Hvec}{\mbox{\boldmath{$H$}}}
\newcommand{\Avec}{\mbox{\boldmath{$A$}}}

\newcommand{\eps}{\varepsilon}
\renewcommand{\phi}{\varphi}

\newcommand{\discrete}{{\scriptscriptstyle dis}}

\theoremstyle{plain}

{\theorembodyfont{\rmfamily}

}

\newcommand{\news}{\setcounter{equation}{0}}

\begin{document}

\title{Skyrmions and spin waves in frustrated ferromagnets at low applied magnetic field}
\author{Martin Speight\thanks{E-mail: {\tt j.m.speight@leeds.ac.uk}} and Thomas Winyard\thanks{E-mail: {\tt t.winyard@leeds.ac.uk}}\\
School of Mathematics, University of Leeds\\
Leeds LS2 9JT, England}

\maketitle

\begin{abstract}
A continuum model of frustrated ferromagnets is analyzed in detail in the regime of low applied magnetic field, $H_0<1/4$, where the ground state is a spatially varying conical spiral. By changing variables to a corotating spin field, the model is reformulated as a gauged sigma model in a fixed background gauge, allowing the construction of stable isolated Skyrmions, and stable multi-Skyrmion clusters, which approach the conical ground state at spatial infinity. Owing to the spatial anisotropy induced by the ground state, these Skyrmions exhibit only discrete symmetries, and are of neither N\'eel nor Bloch type. These Skyrmions are continuously connected to the more familar solutions in the high field regime ($H_0>1/4$), acquiring axial symmetry in the limit $H_0\ra1/4$. The propagation of small amplitude spin waves through the conical ground state is also analyzed and is found to depend strongly on both $H_0$ and propagation direction relative to the ground state. In contrast to spin waves in the high field regime ($H_0>1/4$) there is no spectral gap: waves may propagate with any angular frequency.
\end{abstract}

\news

\section{Introduction}\news\label{sec:intro}

Magnetic Skyrmions are the subject of intense experimental and theoretical study both for their potential applications to data storage \cite{fercrosam} and as an experimentally accessible exemplar of topological solitons \cite{muhbinjon,yuonokan}. The basic Heisenberg model of ferromagnets cannot support stable Skyrmions owing to its instability to scaling variations; some extra mechanism is required to evade Derrick's theorem \cite{der} and stabilize against collapse. Generically, this is provided by the Dzyaloshinsky-Moriya interaction \cite{bogyab} which is first order in spatial derivatives and, crucially, can be negative. In magnetic materials with inversion symmetry, the DMI vanishes identically, so an alternative stabilization mechanism is required. One possibility is magnetic {\em frustration}: spins interact ferromagnetically with near neighbours, but antiferromagnetically with more distant spins \cite{yumostok,okuchukaw,leomos}. This has been proposed for a $J_1$-$J_3$ frustrated Heisenberg model on a triangular lattice \cite{okuchukaw,leomos} and a $J_1$-$J_2$-$J_3$ model on a square lattice \cite{linhay}. The experimental search for inversion symmetric magnetic materials exhibiting stable Skyrmions is in its infancy: as far as we are aware, the first such system was reported only very recently \cite{kur}, and remains, so far, the only known example. A thorough understanding of the stabilization mechanism afforded by frustration, and the distinctive properties of the Skyrmions it induces, cannot but help to inform the search for further examples.

In the continuum limit, magnetically frustrated systems are described \cite{linhay} by an energy with terms both quadratic and quartic in derivatives,
\beq\label{caggf}
E(\svec)=\int_\Omega\left( I_1|\Delta\svec|^2+I_2|\nabla\svec|^2-H_0s_3\right)dx_1 dx_2.
\eeq
Here we imagine a thin sample of material occupying a very large region $\Omega$ in the plane $x_3=0$, described by a unit length spin field $\svec(x_1,x_2)$ subject to an applied magnetic field $\Hvec=(0,0,H_0)$. $I_1,I_2$ are real parameters depending on the spin lattice geometry and relative strengths of the competing lattice site interactions. If the long-range spin interaction is anti-ferromagnetic and sufficiently strong in comparison with the short-range ferromagnetic interaction, then $I_1>0$ while $I_2<0$ \cite{linhay}. By choosing length, energy and magnetic field units appropriately we may, and henceforth will, assume that $I_1=-I_2=1/2$. 

If $H_0\geq 1/4$ the ground state of this system (by which we mean the configuration with lowest total energy) is the {\em spin polarized state}, $\svec=(0,0,1)$, with energy density $\EE_{s-p}=-H_0$. Magnetic Skyrmions \cite{linhay,leomos}, and even three-dimensional knot solitons \cite{sut-frustrated}, have been extensively studied in this regime. In this paper, we examine the low field regime $0\leq H_0<1/4$, where it appears Skyrmions have not yet been constructed. There is a good reason for this: when $0\leq H_0<1/4$, the ground state of this system is not the constant field $\svec=\evec_3$, but rather a spatially varying conical spiral field \cite{leomos}. It is useful for us to write down this field in a particular form. For each $\alpha\in\R$ let $R(\alpha)$ denote the $SO(3)$ matrix producing rotation by $\alpha$ about the $x_3$ axis,
\beq
R(\alpha)=\left(\begin{array}{ccc}\cos\alpha & -\sin\alpha & 0 \\ \sin\alpha & \cos\alpha & 0 \\ 0 & 0 & 1\end{array}\right).
\eeq
Then the conical ground state is
\beq
\svec_0(\xvec)=R(\kvec\cdot\xvec)\uvec_0,
\eeq
where $\kvec=(k_1,k_2)$ is any vector of length $k=1/\sqrt{2}$, and $\uvec_0$ is the constant unit vector $\uvec_0=((1-16H_0^2)^{1/2},0,4H_0)$. This has energy density $\EE_0=-(1+16H_0^2)/8<\EE_{s-p}$. The direction of $\kvec$ is arbitrary, but once chosen, this breaks the rotational invariance of the plane $x_3=0$. As we will see, this has strong effects on the system. Henceforth, without loss of generality, we will choose $\kvec=(k,0)$, aligning the conical ground state along the $x_1$ axis. 

``Skyrmions" in the regime $0\leq H_0<1/4$ should approach the conical ground state as $|\xvec|\ra\infty$, not the constant value $\evec_3$. This introduces a technical difficulty: since $\svec$ is not asymptotically constant, its topological degree (which should be $1$, or $-1$, depending on an orientation convention) is undefined. What we seek is a topological defect sitting ``on top of'' the conical ground state, which should obey $|\svec(\xvec)-\svec_0(\xvec)|\ra 0$ as $|\xvec|\ra\infty$ and should minimize the normalized energy 
\beq\label{Eofs}
E=\int_{\R^2}(\EE-\EE_0)dx_1dx_2=\int_{\R^2}\left(\frac12|\Delta\svec|^2-\frac12|\nabla\svec|^2+\frac18|\svec-4H_0\evec_3|^2\right)dx_1dx_2
\eeq
 among all fields in its homotopy class.  Determining the homotopy class and implementing the boundary conditions are, at first sight, daunting problems. Both are, in fact, easily solved by switching to a corotating reference frame for the spin field. That is, we define $\uvec:\R^2\ra S^2\subset\R^3$ such that
\beq
\svec(\xvec)=R(kx_1)\uvec(\xvec).
\eeq
and demand that $\uvec(\infty)=\uvec_0$. Then the homotopy class of $\svec$ is determined by the topological degree of $\uvec$
\beq
n:=\frac{1}{4\pi}\int_{\R^2}\left(\uvec\times\frac{\cd\uvec}{\cd x_1}\right)\cdot\frac{\cd\uvec}{\cd x_2}\, dx_1 dx_2 
\eeq
which is well defined and integer valued. Although $\svec:\R^2\ra S^2$ does not extend continuously to the one-point compactification $\R^2\cup\{\infty\}$, and so does not have a well-defined topological degree, one can still compute its total topological charge:
\bea
\frac{1}{4\pi}\int_{\R^2}\left(\svec\times\frac{\cd\svec}{\cd x_1}\right)\cdot\frac{\cd\svec}{\cd x_2}\, dx_1 dx_2 &=&
\frac{1}{4\pi}\int_{\R^2}\left\{\left(\uvec\times\frac{\cd\uvec}{\cd x_1}\right)\cdot\frac{\cd\uvec}{\cd x_2}+k\frac{\cd u_3}{\cd x_2}\right\}\, dx_1 dx_2 \nonumber\\
&=&n+\frac{k}{4\pi^2}\int_{\R^2}\nabla\cdot(0,u_3)\, dx_1 dx_2=n
\eea
by the Divergence Theorem and the boundary condition $u_3(\infty)=4H_0$. So the topological degree $n$ of $\uvec$ can also be interpreted as the total
topological charge of $\svec$. 

We shall call a minimizer of $E$ with $n=1$ ($n=-1$) a Skyrmion (anti-Skyrmion). To find such minimizers, we rewrite $E$ in terms of the new field $\uvec$ and minimize over $\uvec$. Note that our setup is crucially different from 
\cite{leomonloubog,rybborblu}, which studied Skyrmions in the low field regime of an unfrustrated system with DM term. In that paper, the precession direction of $\svec$ is {\em parallel} to the applied field, so that on each plane of constant $x_3$ the field $\svec$ obeys a standard Skyrmion boundary condition
($\svec(x_1,x_2,x_3)\ra\svec_0(x_3)$ as $|(x_1,x_2)|\ra\infty$). In our system, the ground state propagation vector is {\em orthogonal} to the applied magnetic field.

The rest of this paper is structured as follows. In section \ref{sec:sigma} we rewrite $E$ as a functional of $\uvec$ and reinterpret the system as a gauged sigma model, with fixed gauge field. We derive the Euler-Lagrange equations and prove that $E(\uvec)\geq 0$ for all fields satisfying appropriate boundary conditions. In section
\ref{sec:numsky}, we use a gradient descent method to numerically minimize $E(\uvec)$, and hence construct $n$-Skyrmions for $n=1,2,\ldots,8$ and various $H_0$. The orientation $\kvec=(k,0)$ of the ground state breaks the rotational symmetry of the system so that these Skyrmions have, at most, a single reflexion symmetry. They are, therefore, very different from the magnetic Skyrmions previously found in the literature. In particular, they cannot meaningfully be classified as being of either N\'eel or Bloch type. In section \ref{sec:spinwaves} we compute the dispersion relation for small amplitude spin waves propagating through the conical ground state, finding strong dependence on $H_0$ and the direction of propagation. Finally, in section \ref{sec:conc} we make some concluding remarks and suggest interesting extensions of this
work.

\section{Reformulation in terms of the corotating field}\news\label{sec:sigma}

Our first task is to rewrite the model's energy functional \eqref{Eofs} in terms of the corotating field $\uvec:=R(kx_1)^{-1}\svec$. This process simplifies considerably 
once we note that
\beq
\cd_{x_1}\svec=R(kx_1)(\cd_{x_1}\uvec+kE_3\uvec), \qquad \cd_{x_2}\svec=R(kx_1)\cd_{x_2}\uvec,\qquad E_3:=\left(\begin{array}{ccc}0 & -1 & 0 \\ 1 & 0 & 0 \\ 0 & 0 & 0\end{array}\right)
\eeq
so $\nabla \svec=R(kx_1) D^A\uvec$ where $D^A=\nabla - \Avec E_3$ and $\Avec=-\kvec=(-k,0)$. This is precisely the gauge covariant derivative of $\uvec$ with respect to the (constant) gauge field $\Avec$, where we have gauged rotations of the target two-sphere about the symmetry axis $\evec_3$. \footnote{We emphasize that this is purely a mathematical device: the gauge field has no physical significance. In particular, it is unrelated to the applied magnetic field. See \cite{barrossch,sch-magnetic} for another interesting application of synthetic gauge fields to ferromagnets.} Computing further, one
sees that
\beq
\Delta\svec=-\nabla^2\svec=-R(kx_1)D^A_iD^A_i\uvec=R(kx_1)\Delta_A\uvec
\eeq
where $\Delta_A=-D^A_iD^A_i$ is the gauge covariant Laplacian. Hence, the energy, as a functional of $\uvec$, assumes the form
\beq
E(\uvec)=\int_{\R^2}\left(\frac12|\Delta_A\uvec|^2-\frac12|D^A\uvec|^2+\frac18|\uvec-4H_0\evec_3|^2\right)dx_1dx_2.
\eeq
The solutions we seek are critical points of $E(\uvec)$. That is, given any smooth variation $\uvec_t$ of $\uvec=\uvec_t|_{t=0}$ of compact support, they must satisfy
\beq
\frac{d\: }{dt}\bigg|_{t=0}E(\uvec_t)=0.
\eeq
Let $\epsvec=\cd_t\uvec_t|_{t=0}$, and note that $\epsvec(x_1,x_2)\cdot\uvec(x_1,x_2)\equiv 0$ since $|\uvec_t(x_1,x_2)|^2\equiv 1$. Hence
\bea
\frac{d\: }{dt}\bigg|_{t=0}E(\uvec_t)&=&\int_{\R^2}\left(\Delta_A\uvec\cdot\Delta_A\epsvec-D^A\uvec\cdot D^A\epsvec+\frac14(\uvec-4H_0\evec_3)\cdot\epsvec\right)dx_1dx_2\nonumber \\
&=&\int_{\R^2}\epsvec\cdot\left(\Delta_A^2\uvec-\Delta_A\uvec-H_0\evec_3\right)
dx_1dx_2
\eea
where we have used the facts that $\Delta_A=(D^A)^\dagger D^A=\Delta_A^\dagger$ (where $\dagger$ denotes $L^2$ adjoint), and that $\epsvec$ has compact support. This integral must vanish for any choice of $\epsvec:\R^2\ra\R^3$ pointwise orthogonal to $\uvec$. Hence, $\uvec$ satisfies
\beq\label{sayo}
P_\uvecs\left(\Delta_A^2\uvec-\Delta_A\uvec-H_0\evec_3\right)=0,
\eeq
where $P_\uvecs:\R^3\ra T_\uvecs S^2$ denotes orthogonal projection,
\beq
P_\uvecs(\vvec):=\vvec-(\uvec\cdot\vvec)\uvec.
\eeq
Equation \eqref{sayo} is the Euler-Lagrange equation for $E(\uvec)$.
Of course, it coincides (after the substitution $\svec=R(kx_1)\uvec$) with the Euler-Lagrange equation derived similarly from \eqref{Eofs} by varying $\svec$.

We seek to solve \eqref{sayo} numerically by gradient descent for the functional $E(\uvec)$. Before doing so, we should check that $E$ is bounded below: owing to the negative second term, this is not immediately clear. So, let $\Omega$ be a bounded region of the plane $x_3=0$ with boundary $\cd\Omega$, and $\uvec:\R^2\ra S^2$. Then,
the energy of the field over the region $\Omega$ is
\bea
E_\Omega(\uvec)&=&\frac12\int_\Omega\left(|\Delta_A\uvec|^2-|D^A\uvec|^2+\frac14|\uvec-4H_0\evec_3|^2\right)dx_1dx_2\nonumber\\
&=&\frac12\int_\Omega\left\{|\Delta_A{\uvec}-\frac12(\uvec-4H_0\evec_3)|^2+(\uvec-4H_0\evec_3)\cdot\Delta_A\uvec-|D^A\uvec|^2\right\}dx_1dx_2.
\eea
An application of Stokes's Theorem yields the general identity
\beq\label{agj}
\int_\Omega\vvec\cdot\Delta_A\uvec\, dx_1dx_2=\int_{\Omega}D^A_i\vvec\cdot D^A_i\uvec\, dx_1dx_2-\oint_{\cd\Omega}\vvec\cdot D^A_i\uvec\, n_i ds
\eeq
where $\nvec=(n_1,n_2)$ is the outward unit normal to the closed curve $\cd\Omega$, and $s$ is an arclength parameter. Exploiting \eqref{agj} in the case 
$\vvec=\uvec-4H_0\evec_3$, we see that
\beq
E_\Omega(\uvec)=\frac12\int_\Omega|\Delta_A\uvec-\frac12(\uvec-4H_0\evec_3)|^2 dx_1dx_2-\oint_{\cd\Omega}(\uvec-4H_0\evec_3)\cdot D^A_i\uvec\, n_i ds.
\eeq
Since $|\uvec|^2\equiv 1$ and $E_3$ is skew, $\uvec\cdot D^A_i\uvec=0$. Furthermore, $\evec_3\cdot E_3\uvec=-\uvec\cdot E_3\evec_3=0$. Hence
\beq
E_\Omega(\uvec)=\frac12\int_\Omega|\Delta_A\uvec-\frac12(\uvec-4H_0\evec_3)|^2 dx_1dx_2+4H_0\oint_{\cd\Omega}\frac{\cd u_3}{\cd x_i}\, n_i ds.
\eeq
Consider now the case where $\Omega$ is the disk where $|\xvec|\leq R$ and $\uvec(\xvec)\ra\uvec_0$ as $|\xvec|\ra\infty$, so that $\cd_r u_3\ra 0$ faster than $1/r$. Then the energy of the field on the whole plane is
\beq
E(\uvec)=\lim_{R\ra\infty}E_\Omega(\uvec)=\frac12\int_{\R^2}|\Delta_A\uvec-\frac12(\uvec-4H_0\evec_3)|^2dx_1dx_2\geq 0.
\eeq

So every field decaying to $\uvec_0$ sufficiently fast (in fact, every field with $u_3$ asymptotically constant) has non-negative energy. Furthermore, $E(\uvec)=0$
 if and only if
\beq\label{asgijo}
\Delta_A\vvec-\frac12\vvec=0
\eeq
where, once again, $\vvec=\uvec-4H_0\evec_0$, and we have observed that $\Delta_A\evec_3=0$. Taking the scalar product of \eqref{asgijo} with $\evec_3$, we see that
$\Delta v_3=\frac12 v_3$ and the boundary condition implies that $v_3\ra 0$ at spatial infinity. The Laplacian on $\R^2$ has no decaying eigenfunctions, so $v_3\equiv 0$. Hence $\vvec=(1-16H_0^2)^{1/2}(\cos\chi,\sin\chi,0)$ for some smooth function $\chi:\R^2\ra\R$ tending to $0$ at spatial infinity. Substituting this into \eqref{asgijo}
yields a vector-valued PDE 
\beq
\Delta\chi(-\sin\chi,\cos\chi,0)+(|\d\chi|^2+2k\chi_x+k^2)(\cos\chi,\sin\chi,0)=\frac12(\cos\chi,\sin\chi,0),
\eeq
whose $(-\sin\chi,\cos\chi,0)$ component implies $\Delta\chi=0$ and hence, by virtue of the boundary condition, $\chi\equiv 0$, that is, $\uvec(\xvec)=\uvec_0$.
Hence, provided $\uvec\ra\uvec_0$ as $|\xvec|\ra \infty$, $E(\uvec)\geq 0$ with equality if and only if $\uvec$ is the conical ground state.

This result is subtle. We have {\em not} shown that the energy {\em density} of our model is non-negative, only that the total energy is non-negative for all fields satisfying an appropriate boundary condition. 
Since $E(\uvec)$ coincides with the alternative functional 
\beq
\wt{E}(\uvec)=\frac12\int_{\R^2}|\Delta_A\uvec-\frac12(\uvec-4H_0\evec_3)|^2dx_1dx_2,
\eeq
the Euler-Lagrange equation for $\wt{E}$ should coincide with \eqref{sayo}. It is a straightforward exercise to verify that this is true.
The general strategy for obtaining the energy bound above was suggested by Harland's work on the model \eqref{caggf} in the high field regime $H_0>1/4$, where the ground state is uniform $\svec(\xvec)=\evec_3$. In this setting, Harland \cite{har-mag} established a topological lower energy bound of the form $E(\svec)\geq C(H_0)|n|$ where $n$ is the topological degree of $\svec$ and $C(H_0)$ is a positive constant depending on $H_0$, with $C(1/4)=0$. It seems likely that a similar bound holds also for $E(\uvec)$ when $0<H_0<1/4$, but we have been unable to prove this. As we will see, our numerical results are certainly consistent with such a bound.

\section{Isolated Skyrmions and Skyrmion clusters}\news\label{sec:numsky}

\begin{figure}
\includegraphics[width=1.0\linewidth]{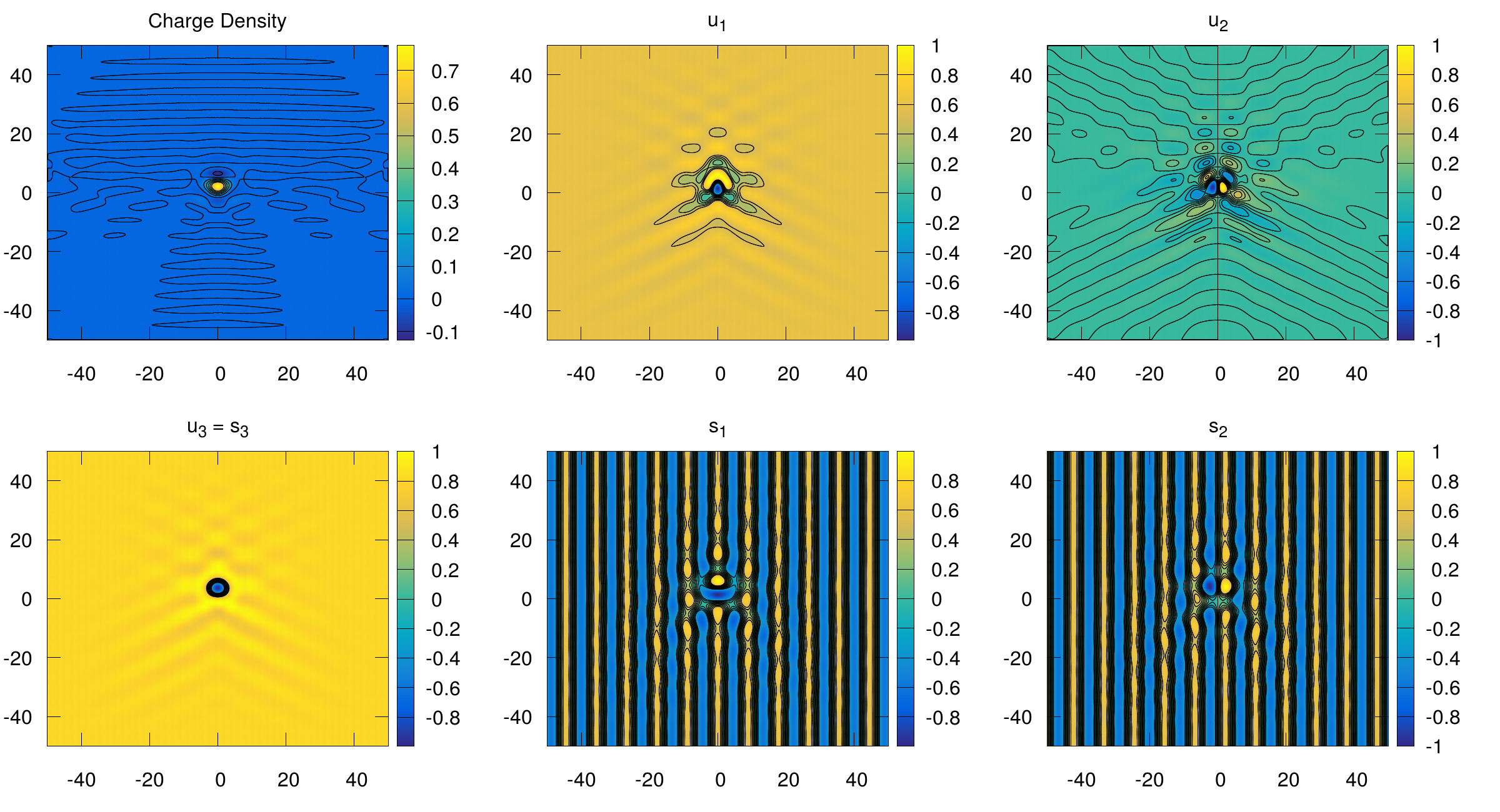}
 \caption{\label{Fig:charge1}
Contour plots of a charge one Skyrmion in external field $H_0 = 0.2$. We have plotted the topological charge density, the components  $u_1$,$u_2$ and $u_3$ of the corotating field and the components $s_3$, $s_1$ and $s_2$ of the original magnetisation field. Note that $s_3\equiv u_3$ by definition.}
\end{figure}

In this section we present numerically generated energy minimizers of degrees $n=1,2\ldots,8$, obtained by a gradient descent method. We first discretize the model, placing it on a regular $N_1\times N_2$ grid with spacing $h>0$ (typical values are {$N_1 = N_2 = 500$ and $h = 0.2$}), and replacing $\cd_{x_1}$, $\cd_{x_2}$ and $\Delta$ by standard difference operators. This yields a discrete approximant $E_{\discrete}$ to the functional $E(\uvec)$, which we may regard as a function $E_\discrete:\CC\ra \R$, where
the discretized configuration space is the manifold $\CC=(S^2)^{N_1N_2}\subset \R^{3N_1N_2}$. We now seek local minima of $E_\discrete$ subject to the constraint that
$\uvec=\uvec_0$ on the edge of the computational grid. To find such minima we use arrested Newton flow: we solve Newton's equation for the motion of a notional ``particle" in $\CC$ subject to the potential $E_\discrete$,
\beq
\ddot\uvec=-\grad E_\discrete(\uvec),
\eeq
starting at some initial guess $\uvec(0)$ with $\dot{\uvec}(0)=0$. This flow naturally begins to run ``downhill". After each time step $t\mapsto t+\delta t$, we check whether $E_\discrete(t+\delta_t)>E_\discrete(t)$. If so, we set $\dot{\uvec}(t+\delta t)=0$ and restart the flow. The flow terminates at an acceptable
approximant to an energy minimizer when every component of $\grad E_\discrete (\uvec)$ is zero to within a pre-assigned tolerance (we used {$10^{-4}$}). This scheme is robust, simple and much faster than simple gradient flow (solving $\dot\uvec=-\grad E_\discrete$). An estimate of the numerical error in $E_\discrete$ can be obtained by computing the discretized topological charge
\beq
n_\discrete=\frac{h^2}{4\pi}\sum_{i,j} \left(\uvec_{i,j}\times\cd_1\uvec_{i,j}\right)\cdot\cd_2\uvec_{i,j} ,
\eeq
where the derivatives are approximated using a fourth order central finite difference approximation
($\cd_1\uvec_{i,j} = {(-\uvec_{i+2,j} + 8 \uvec_{i+1,j} - 8\uvec_{i-1,j} + \uvec_{i-2,j})}/{(12h)}$,\quad
$\cd_2\uvec_{i,j}$ defined similarly), 
and comparing it with the integer $n$. In all the simulations reported here $|n_\discrete-n|<{10^{-4}}$. As with any gradient descent method, the final field is a {\em local} minimizer of $E_\discrete$. We must start the algorithm with several different choices of initial field to be confident of finding the global energy minimizer in a given homotopy class of fields (that is, for a given topological charge $n$).

\begin{figure}
\includegraphics[width=1.0\linewidth]{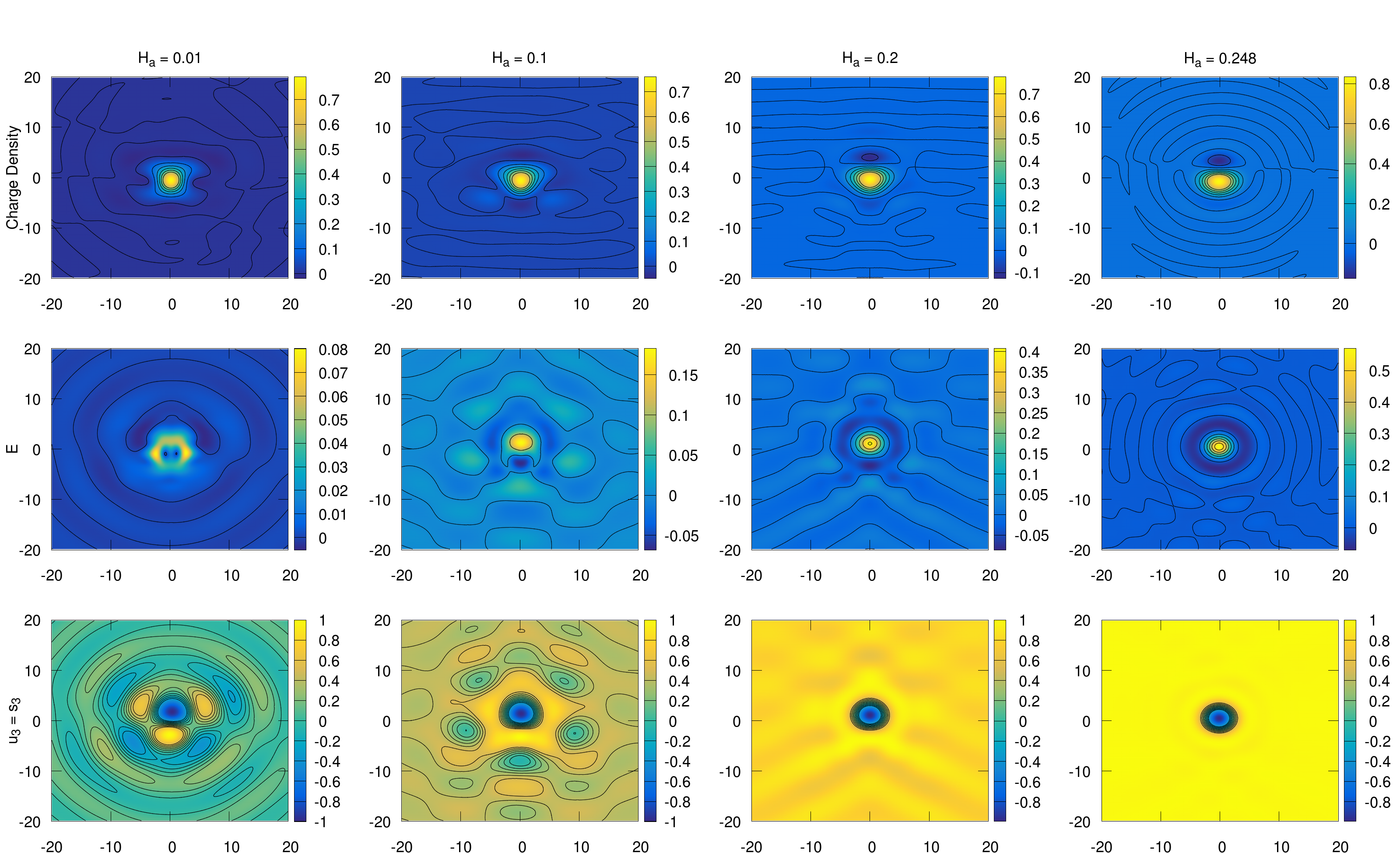}
 \caption{\label{Fig:charge1s}
Contour plots for charge one Skyrmions in external fields $H_0 = 0.01$,$0.1$,$0.2$ and $0.248$ from left to right. The top row depicts the topological charge density of $\uvec$, the middle row is energy density, while the bottom row shows $s_3\equiv u_3$.}
\end{figure}

The charge $n=1$ energy minimizer for applied field $H_0=0.2$ is presented in figure \ref{Fig:charge1}. Two striking features are evident: this soliton has very little symmetry, and is quite large, with a long range tail exhibiting nontrivial angular dependence. The core of the soliton occupies a region of width $\approx$ {9}, which is comparable to the spatial period of the conical ground state, $2\pi/k\approx 8.89$. Outside this core, the field ripples around $\uvec_0$ out to quite a long range. To accommodate this long tail, we need a large computational grid, and to resolve its spatial structure, the grid must be fairly finely discretized. 

The lack of symmetry is inevitable. The choice of conical ground state orientation $\kvec=(k,0)$ and boundary value $\uvec_0$ breaks the $O(2)\times O(2)$ symmetry of the model to only two reflexion symmetries:
\bea
\uvec(x_1,x_2)&\mapsto& (u_1(-x_1,x_2),-u_2(-x_1,x_2),u_3(-x_1,x_2))\label{sym1}\\
\uvec(x_1,x_2)&\mapsto& \uvec(x_1,-x_2).\label{sym2}
\eea
Of these, \eqref{sym1} preserves topological charge $n$, while \eqref{sym2} maps $n$ to $-n$. Hence, static solutions with $n\neq 0$ will, at most, be symmetric under the single reflexion symmetry \eqref{sym1}. We may use \eqref{sym2} to obtain charge $-n$ Skyrmions from charge $n$ Skyrmions, and hence, without loss of generality, consider only $n>0$. 

{
Translation symmetry is retained, however, which means that Skyrmions do not have an optimal position relative to the conical ground state. We can freely translate the Skyrmion in the co-rotating frame, $\uvec(x_1,x_2)\mapsto\uvec(x_1-\mu_1,x_2-\mu_2)$, which corresponds to the Skyrmion's spin field undergoing translation relative to the ground state coupled with a rotation in the target space,
\begin{equation}
\boldsymbol{s}(x_1, x_2) \mapsto R(k\mu_1)\boldsymbol{s}(x_1-\mu_1, x_2-\mu_2).
\end{equation}
}

{Figure \ref{Fig:charge1s} shows the dependence of the charge $1$ Skyrmion on the applied field $H_0$. 
It is important to remember that the boundary condition changes as $H_0$ changes, hence the form of $s_3$ changes dramatically with $H_0$. Note, however, that the size of the soliton core does not vary significantly. These plots also show the energy density of the solutions, and it is interesting to note that there are pockets of {\em negative} energy density, meaning regions where the Skyrmion has lower energy density than the conical ground state. As discussed in section \ref{sec:sigma} this does not contradict our theorem that the total energy is non-negative.}

\begin{figure}
\includegraphics[width=1.0\linewidth]{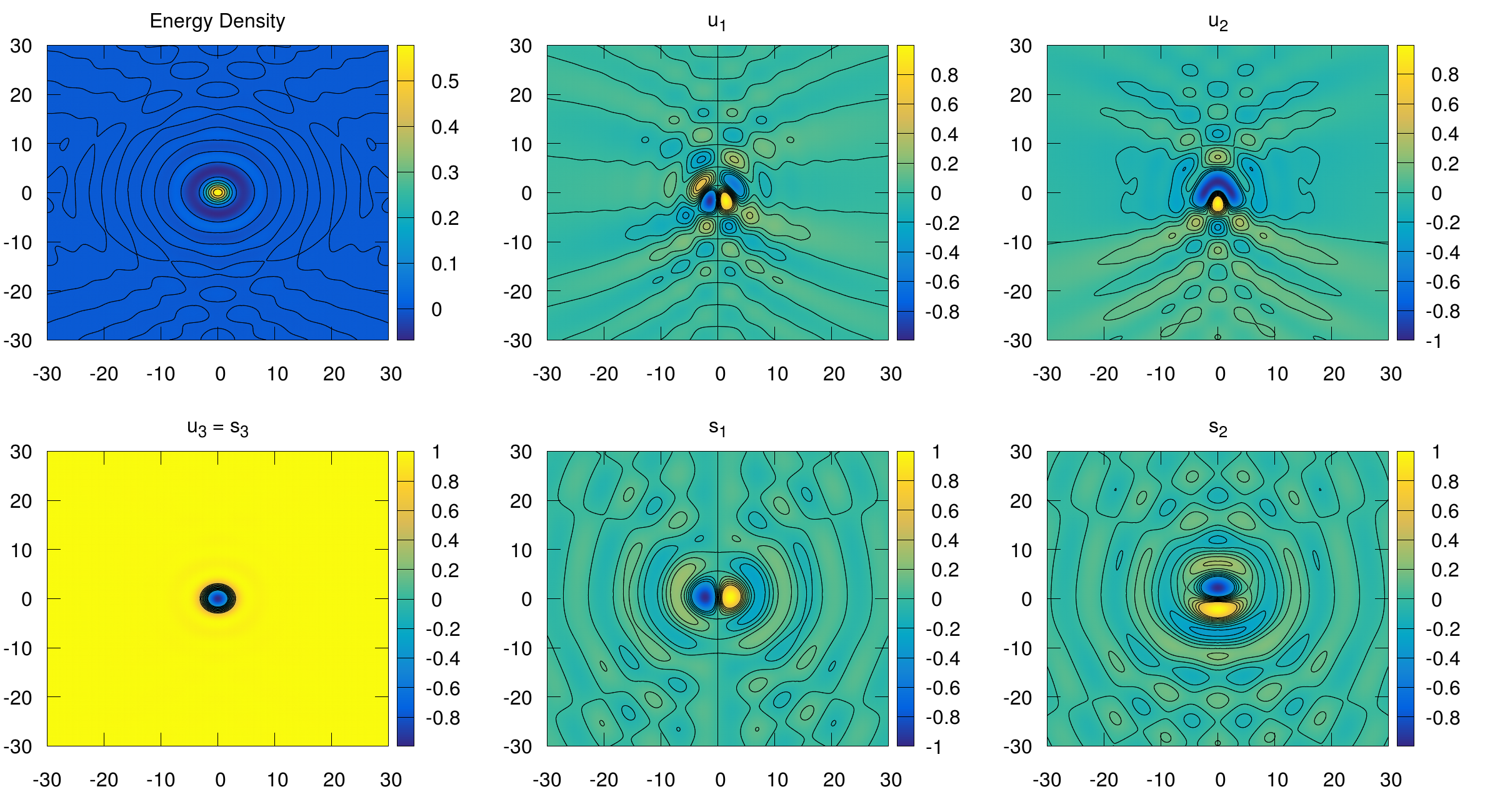}
 \caption{\label{Fig:charge1critical}
{Contour plots of a charge one Skyrmion with critical external field $H_0 = 1/4$. We have plotted the normalized energy density, the components  $u_1$,$u_2$ and $u_3$ of the corotating field and the components $s_3$, $s_1$ and $s_2$ of the original magnetisation field. Note that $s_3\equiv u_3$ by definition and that the original spin field $\svec$ exhibits approximate axial symmetry.}}
\end{figure}

\begin{figure}
\includegraphics[width=1.0\linewidth]{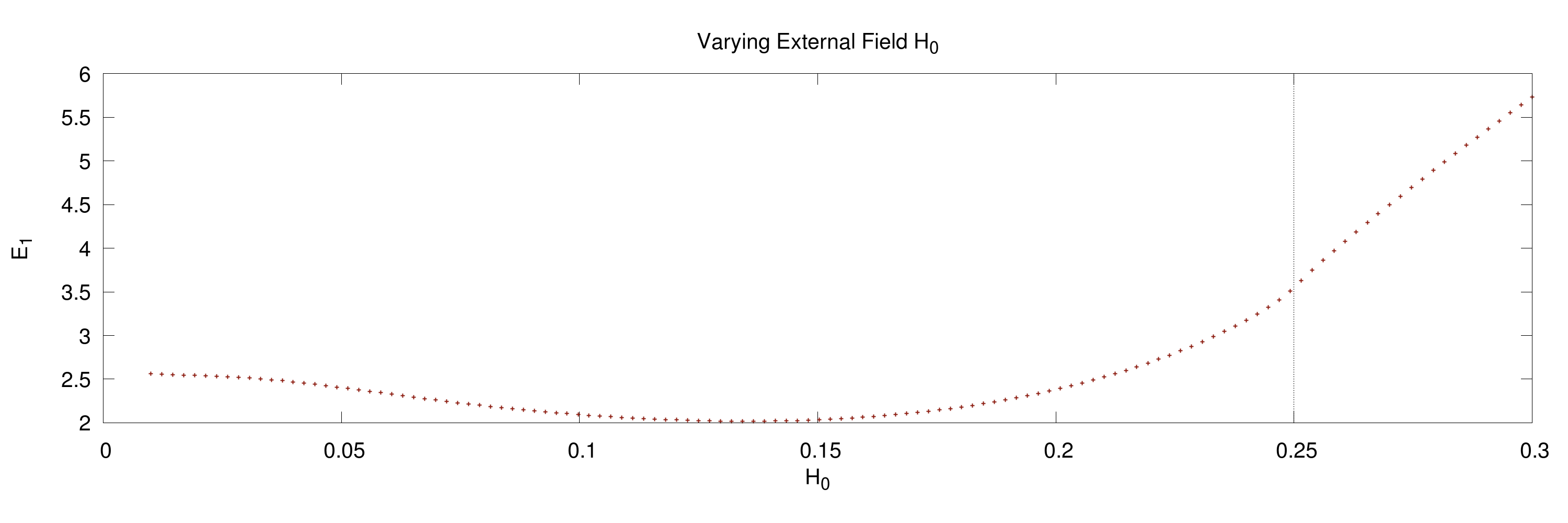}
 \caption{\label{Fig:varyHa}
{Plot of $E_1$, the normalized energy of a charge one Skyrmion, as a function of the external field strength $H_0$. The critical field value $H_0 = 1/4$ is marked with a dashed line where the solutions change from the familiar radial Skyrmions ($H_0>1/4$) to the nematic Skyrmions discussed in this paper ($H_0<1/4$). Note that the boundary conditions and the normalizing energy density are piecewise functions of $H_0$ given in \eqref{Eq:piece}.} }
\end{figure}

{ Let us denote by $E_n$ the energy $E(\uvec)$ of the lowest energy charge $n$ solution, and recall that this is, by definition, the
energy excess of $\uvec$ over the conical ground state for $H_0 < 1/4$, or the polarized ground state $\uvec=\evec_3$ for $H_0\geq 1/4$. As shown in section \ref{sec:sigma}, 
$E_n>0$ for all $n\neq 0$. Figure \ref{Fig:varyHa} presents a plot of $E_1$ as a function of $H_0$, showing that the energy required for a single Skyrmion above the ground state is less for all $H_0 < 1/4$ than it is for $H_0>1/4$, suggesting Skyrmions in low field are more stable than their radially symmetric high field counterparts. Note that $E_1(H_0)$ attains a minimum at 
$H_0 \approx 0.135$,  and remains bounded as $H_0\ra 0^+$. It also remains bounded as it approaches the critical external field value $H_0 = 1/4$ and passes smoothly through this value without the Skyrmion core size changing significantly. It has been suggested elsewhere { \cite{linhay} }that a Skyrmion cannot exist for $H_0 = 1/4$, but we find that there is nothing preventing this. Figure \ref{Fig:charge1critical} presents the charge 1 solution for $H_0 = 1/4$, which appears to have axial symmetry. Note that our formulation
of the energy minimization problem works for all values of $H_0$ with the caveat that the boundary value,  normalizing energy density and background gauge field are modified to the following piecewise functions of $H_0$,
\begin{equation}
[\uvec_0,\EE_0,\Avec]=\left\{\begin{array}{cc} [((1-16H_0^2)^{1/2},0,4H_0),-(1+16H_0^2)/8,(k,0)], & H_0 \leq \frac{1}{4}, \\ {[(0,0,1),-H_0,(0,0)]}, & H_0 \geq \frac{1}{4}. \end{array}\right. 
\label{Eq:piece}
\end{equation}}

\begin{figure}
\includegraphics[width=1.0\linewidth,trim={4cm 14cm 0cm 8.6cm},clip]{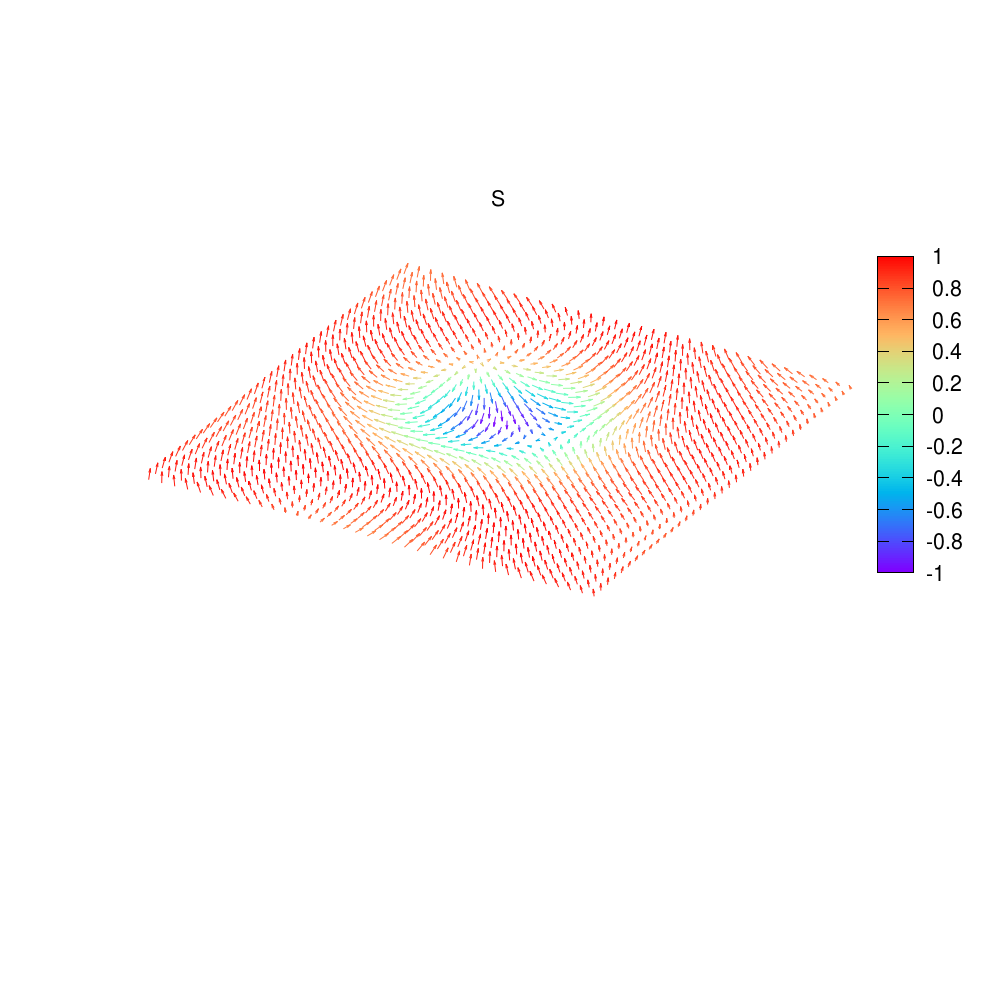}
 \caption{\label{Fig:arrows}
Arrows depicting the spin field $\boldsymbol{s}$ of the charge one Skyrmion in external field $\boldsymbol{H} = (0,0,0.2)$. The colour indicates the value of $s_3$. Note that the Skyrmion is neither N\'eel nor Bloch type, due to the anisotropy present. }
\end{figure}

{We have plotted the field $\boldsymbol{s}$ in figure \ref{Fig:arrows} for $H_0 = 0.2$, which makes it clear that these Skyrmions can be thought of as neither N\'eel nor Bloch type. The form of the field $\boldsymbol{s}$ is clearly $H_0$ dependent and strongly affected by the anisotropy of the background.}

\begin{figure}
\includegraphics[width=1.0\linewidth]{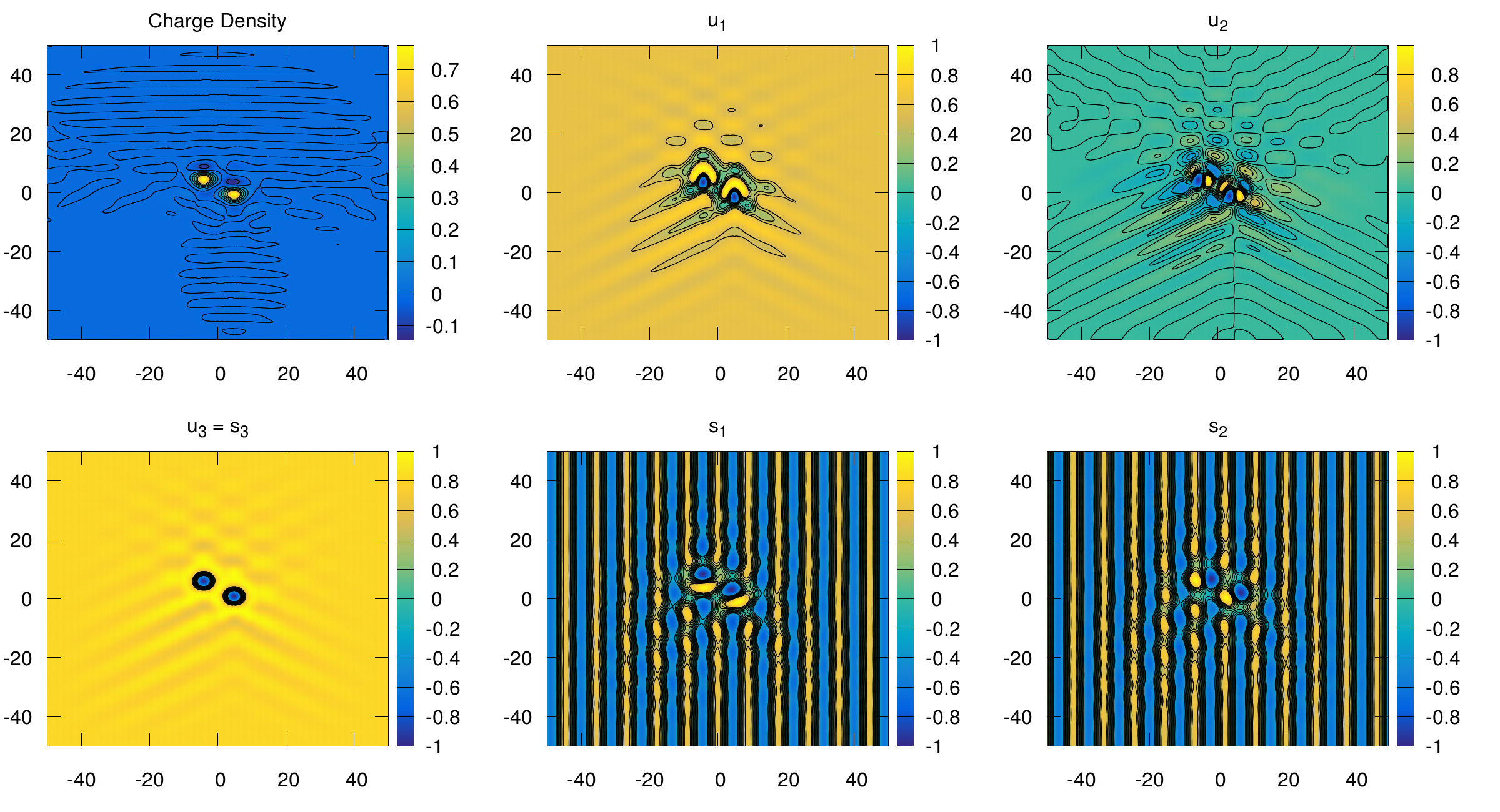}
 \caption{\label{Fig:charge2}
Contour plot of a charge two energy minimizer in external field $H_0 = 0.2$. We have plotted the topological charge density and components $u_1$, $u_2$ and $u_3$, of the corotating fields $\uvec$ and the original magnetization fields $s_1$ and $s_2$ (recall $s_3\equiv u_3$).}
\end{figure}

{The charge $n=2$ energy minimizer for applied field $H_0=0.2$ is presented in figure \ref{Fig:charge2}. The minimizer is a bound state of two individual Skyrmions, that is, $E_2 < 2 E_1$, and the field resembles a superposition of two unit Skyrmions sitting a short distance from one another.
Note that the direction between the two Skyrmion centres is not a free parameter of this solution, but is determined by the orientation of the
background 
conical ground state. }

\begin{figure}
\includegraphics[width=1.0\linewidth]{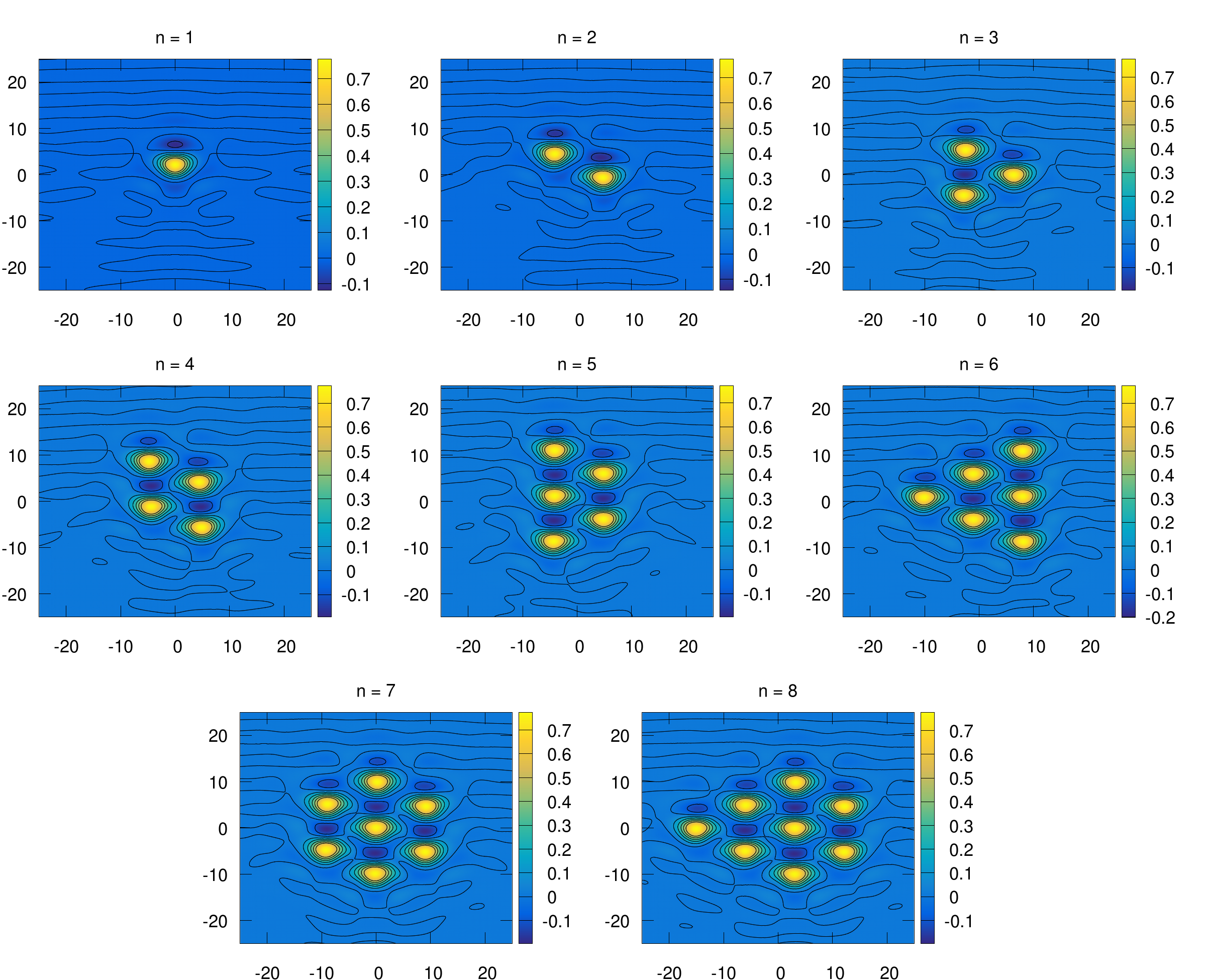}
 \caption{\label{Fig:charge18}
Contour plots of the topological charge density for energy minimizers of charge $n=1,2,\ldots,8$ in external field $H_0 = 0.2$. The energy minimizers resemble clusters of single Skyrmions.}
\end{figure}

\begin{figure}
\includegraphics[width=\linewidth]{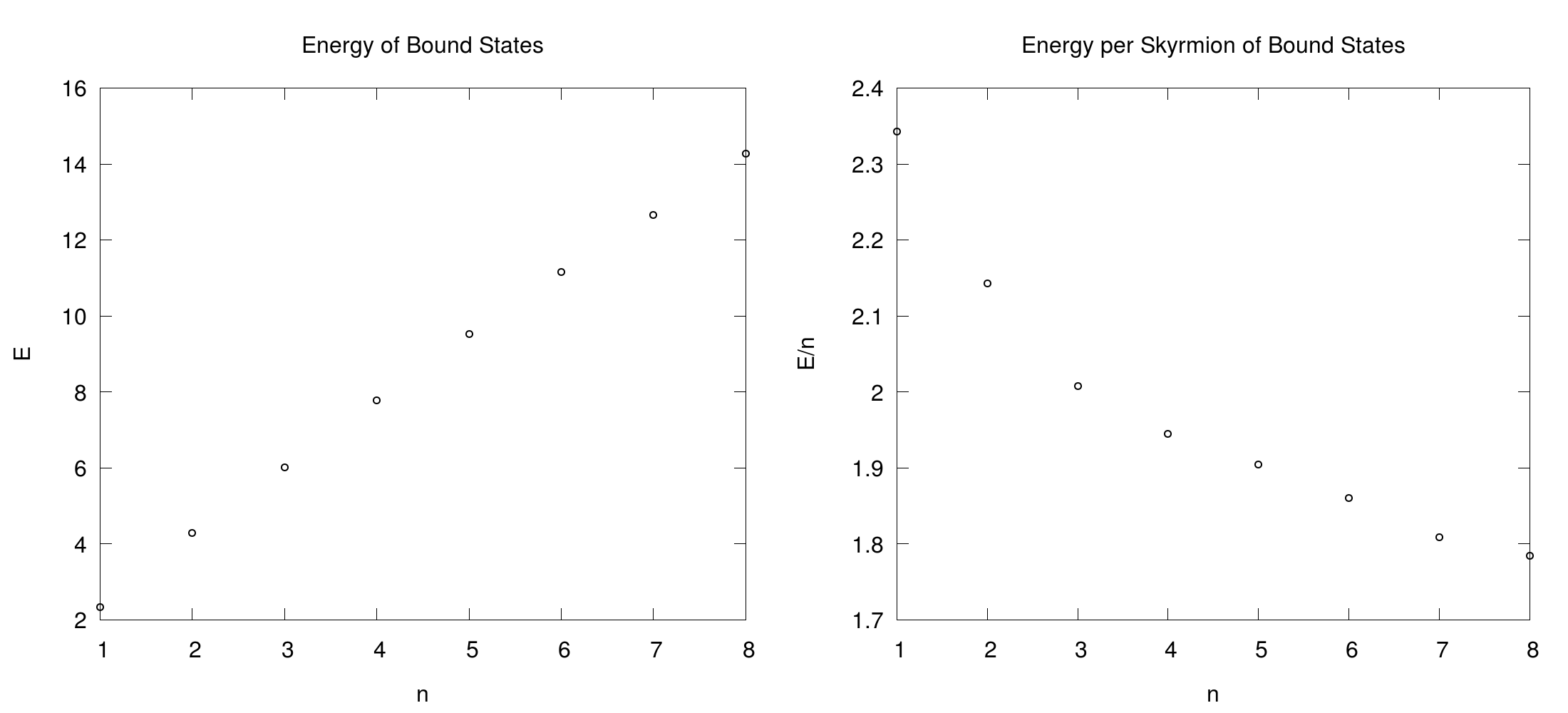}
 \caption{\label{Fig:energies}
Left: a plot of the energy $E_n$ of the charge $n$ minimizers for $n=1,2,\ldots,8$. Note the approximately linear energy growth. Right: energy per unit  Skyrmion ($E_n/n$), which one should note decreases with $n$, showing that $n$-Skyrmions are stable against fission into lower charge subclusters.}
\end{figure}

{For higher $n$ we find a growing number of local energy minima, corresponding to bound states where $n$ individual Skyrmions position themselves favourably relative to each others' tails. Figure \ref{Fig:charge18} depicts a selection of these. Such bound states, with given $n$, tend to have nearly degenerate energy. In figure \ref{Fig:energies} we plot $E_n$ and $E_n/n$ versus $n$ for $1\leq n\leq 8$ and $H_0=0.2$. These data suggest roughly linear energy growth with topological charge, and stability of charge $n$ Skyrmions against fission into lower charge clusters.  }

\section{Spin waves}\news\label{sec:spinwaves}

In this section we study the propagation of small amplitude spin waves through the ground state. Time evolution of the spin field $\svec$, in the absence of applied electric fields and neglecting damping, is governed by the equation
\beq
\label{sy}
{\svec}_{t}=-\svec\times\grad E(\svec)=-\svec\times(\Delta^2\svec-\Delta\svec-H_0\evec_3),
\eeq
where we have set the gyromagnetic ratio to unity by an appropriate choice of time unit. If $H_0>1/4$ the ground state is spin polarized, $\svec=\evec_3$, and small amplitude spin waves take the form $\svec(\xvec,t)=\evec_3+\epsvec(\xvec,t)$ where $\epsvec$ is small and, to leading order, is orthogonal to $\evec_3$. The evolution of $\epsvec$ is governed by the linearization of \eqref{sy} about $\svec=\evec_3$,
\beq
\label{sylmc}
\epsvec_t=-\evec_3\times(\Delta^2\epsvec-\Delta\epsvec+H_0\epsvec),
\eeq
where $\epsvec=(\eps_1,\eps_2,0)$. This coupled pair of PDEs for $(\eps_1,\eps_2)$ can be conveniently written as a single complex valued PDE for
$\eps=\eps_1+i\eps_2$:
\beq\label{sylmcplx}
\eps_t=-i(\Delta^2\eps-\Delta\eps+H_0\eps).
\eeq
Equation \eqref{sylmcplx} supports travelling waves of the form $\eps(\xvec,t)=\exp(i(\mvec\cdot\xvec-\omega t))$ provided the frequency $\omega$ and wavevector $\mvec=(m_1,m_2)$ satisfy the dispersion relation
\beq\label{hfdisp}
\omega=|\mvec|^4-|\mvec|^2+H_0=\left(|\mvec|^2-\frac12\right)^2+\left(H_0-\frac14\right).
\eeq
Note that the dispersion relation is isotropic ($\omega$ depends only on $|\mvec|$), taking the form of a ``Mexican hat" which approaches and touches the plane $\omega=0$ in the limit $H_0\ra 1/4$ from above. Hence, there is a spectral gap, $\omega\geq H_0-1/4$.

We seek to repeat this analysis in the regime $0<H_0<1/4$, where the ground state is conical. The first task is to rewrite the evolution equation \eqref{sy} in terms of the corotating field $\uvec(\xvec,t)=R(kx_1)^{-1}\svec(\xvec,t)$. One finds that
\beq\label{lifpsi}
\uvec_t=-\uvec\times(\Delta_A^2\uvec-\Delta_A\uvec-H_0\evec_3).
\eeq
We seek small amplitude travelling wave solutions to this equation about the conical ground state, $\uvec=\uvec_0=(a,0,b)$ with $b=4H_0$ and $a=\sqrt{1-b^2}$. These take the form $\uvec(\xvec,t)=\uvec_0+\epsvec(\xvec,t)$ where $\epsvec$ is valued in $T_{\uvecs_0} S^2$ (is orthogonal to $\uvec_0$) and satisfies the linearization of \eqref{lifpsi} about the static solution $\uvec=\uvec_0$,
\beq\label{liflin}
\epsvec_t=-\uvec_0\times(\Delta_A^2\epsvec-\Delta_A\epsvec+\frac14\epsvec).
\eeq
It is convenient to use the basis 
\beq
\wvec_1:=(0,1,0),\quad \wvec_2:=(-b,0,a),
\eeq
for $T_{\uvecs_0}S^2$, noting that $[\wvec_1,\wvec_2,\uvec_0]$ is an oriented orthonormal frame for $\R^3$. Then $\epsvec(\xvec,t)=\eps_1(\xvec,t)\wvec_1+\eps_2(\xvec,t)\wvec_2$, and \eqref{liflin} is equivalent to the linear PDE system
\beq\label{lifmat}
\eps_t=-J\left\{\Delta^2\eps-16kH_0J\Delta\eps_x+K(\Delta\eps-2\eps_{xx})-\Delta\eps-\frac14K\eps+\frac14\eps\right\},
\eeq
for the two-vector $\eps=(\eps_1,\eps_2)$, where $J,K$ denote the matrices
\beq\label{JKdef}
J:=\left[\begin{array}{cc}0&-1\\1&0\end{array}\right],\quad
K:=\left[\begin{array}{cc}1&0\\0&16H_0^2\end{array}\right].
\eeq

We choose and fix a wavevector $\pvec=(p_1,p_2)$, and seek solutions of the form
\beq\label{skilif}
\eps(\xvec,t)={\rm Re}(v\exp(i(\pvec\cdot\xvec-\omega t))),
\eeq
where $\omega\in\R$ and $v\in\C^2$ are constants depending on $\pvec$, to be determined. Substituting \eqref{skilif} into \eqref{lifmat}, one sees that $\omega$ must be an
eigenvalue of the complex $2\times 2$ matrix
\beq\label{Omegadef}
\Omega=16kH_0p_1|\pvec|^2\I_2-i\left((|\pvec|^2-\frac12)^2J+(|\pvec|^2+2p_1^2-\frac14)JK\right),
\eeq
and $v$ a corresponding eigenvector. This matrix may be usefully decomposed as $\Omega=:\lambda\I_2+\Omega_0$ where
$\lambda=16kH_0p_1|\pvec|^2$ and
\bea
\Omega_0&=&\left[\begin{array}{cc}0&i\alpha(\pvec)\\-i\beta(\pvec)&0\end{array}\right],\nonumber\\
\alpha(\pvec)&=&(|\pvec|^2-\frac12)^2+16H_0^2(|\pvec|^2+2p_1^2-\frac14),\nonumber\\
\beta(\pvec)&=&(|\pvec|^2-\frac12)^2+|\pvec|^2+2p_1^2-\frac14=|\pvec|^4+2p_1^2.
\eea
Clearly, for all $\pvec$, 
\beq\label{chsbd}
\alpha(\pvec)\geq 16H_0^2\beta(\pvec)\geq 0
\eeq
with equality at the first stage if and only if $|\pvec|^2=1/2$. The eigenvectors of $\Omega_0$ are
\beq
v_\pm=\left[\begin{array}{c}\sqrt{\alpha(\pvec)}\\ \pm i\sqrt{\beta(\pvec)}\end{array}\right],
\eeq
with corresponding eigenvalues $\pm\sqrt{\alpha(\pvec)\beta(\pvec)}$. Hence $\Omega$ also has eigenvectors $v_\pm$, and eigenvalues
\beq\label{worbit}
\omega_\pm(\pvec)=16kH_0p_1|\pvec|^2\pm\sqrt{\alpha(\pvec)\beta(\pvec)}.
\eeq

Ostensibly, then, for each choice of wavevector $\pvec$, we have two distinct spin waves,
\beq
\eps_\pm^{\pvecs}(\xvec,t)=\left[\begin{array}{c}
    \sqrt{\alpha(\pvec)}\cos(\pvec\cdot\xvec-\omega_\pm(\pvec)t)\\ 
\pm \sqrt{\beta(\pvec)}\sin(\pvec\cdot\xvec-\omega_\pm(\pvec)t)\end{array}\right].
\eeq
One should note, however, that $\eps_-^{\pvecs}\equiv\eps_+^{-\pvecs}$, so we have really double-counted the solutions: the general spin wave takes the form
\beq
\epsvec^{\pvecs}(\xvec,t)=\sqrt{\alpha(\pvec)}\cos(\pvec\cdot\xvec-\omega(\pvec)t)\wvec_1+\sqrt{\beta(\pvec)}\sin(\pvec\cdot\xvec-\omega(\pvec)t)\wvec_2,
\eeq
where 
\beq\label{disprel}
\omega(\pvec)=\sqrt{\alpha(\pvec)\beta(\pvec)}+16kH_0p_1|\pvec|^2.
\eeq
 Now, by the AM-GM inequality,
\beq\label{amgm}
\beta(\pvec)\geq 2\sqrt{2}|p_1||\pvec|^2
\eeq
with equality if and only if $|\pvec|^4=2p_1^2$. Hence
\bea
\omega(\pvec)
&\geq& 4H_0\beta(\pvec)+16kH_0p_1|\pvec|^2\qquad\mbox{by \eqref{chsbd}} \nonumber \\
&\geq& 16kH_0(|p_1|+p_1)|\pvec|^2\qquad\mbox{by \eqref{amgm}} \nonumber \\
&\geq& 0
\eea
with equality if and only if $p_1\leq 0$ and $|\pvec|^2=1/2$ or $\beta(\pvec)=0$. That is, $\omega(\pvec)\geq 0$ and $\omega(\pvec)=0$ precisely for the three
wavevectors $\pvec=(0,0)$, $\pvec=(-k/2,\sqrt{3}k/2)$ and $\pvec=(-k/2,-\sqrt{3}k/2)$. The first of these is the zero mode associated with translating the conical
ground state in the $x_1$ direction, but the other two are somewhat mysterious. By continuity, the range of $\omega$ is $[0,\infty)$ so, in contrast to the high field
regime, there is no spectral gap.

Recall that $\epsvec$ describes the spin wave in the corotating frame. The actual spin dynamics is
\beq
\svec(\xvec,t)=R(kx_1)(\uvec_0+\ell\epsvec^{\pvecs}(\xvec,t)+O(\ell^2)),
\eeq
where $\ell>0$ is a small parameter determining the amplitude of the wave. As time evolves, the spin at any fixed position $\xvec$ precesses clockwise around a small ellipse centred on $R(kx_1)\uvec_0$, whose eccentricity depends on $\pvec$ (and $H_0$). This ellipse tends to a flat line in the limit $\pvec\ra (0,0)$ (since $\beta(0,0)=0$ while
$\alpha(0,0)=(1-16H_0^2)/4>0$). 

Clearly, the dispersion relation \eqref{disprel} is anisotropic: $\omega(\pvec)$ depends on the direction of $\pvec$, not just $|\pvec|$.  This is to be expected, since the spatial orientation of the conical ground state (here chosen to be aligned with the $x_1$-axis) breaks the system's spatial isotropy. One should note that the ground state also breaks the reflexion symmetry $(x_1,x_2)\mapsto (-x_1,x_2)$ and, correspondingly, $\omega(-p_1,p_2)\neq \omega(p_1,p_2)$. That is, {\em forward} propagation of
spin waves along the ground state ($p_1>0$) is not equivalent to {\em backward} propagation ($p_1<0$). The difference in frequencies is
\beq\label{hmc}
\omega(p_1,p_2)-\omega(-p_1,p_2)=32kH_0p_1|\pvec|^2. 
\eeq
When interpreting \eqref{hmc} one should bear in mind that $\omega(\pvec)$ describes the frequency of spin waves in a spatially corotating frame. In the limit 
$H_0\ra 1/4^-$,
\beq\label{lfq}
\omega_{1/4^-}=|\pvec|^4+4kp_1|\pvec|^2+2p_1^2=\left(|\pvec+k\evec_1|^2-\frac12\right)^2,
\eeq
whereas in the limit $H_0\ra1/4^+$, \eqref{hfdisp} yields,
\beq\label{hfq}
\omega_{1/4^+}=\left(|\mvec|^2-\frac12\right)^2.
\eeq
The apparent contradiction is resolved once we recognize that \eqref{lfq} describes spin waves about the uniform ground state $\svec=\evec_3$ with respect to the corotating frame $[R(kx_1)\evec_2,-R(kx_1)\evec_1]$ for
$T_{\evecs_3}S^2$, while \eqref{hfq} describes the same spin waves but with respect to the constant frame $[\evec_1,\evec_2]$. For this reason, the definition of
wavevector differs for the two calculations. To match them we must identify $\mvec\equiv\pvec+k\evec_1$ at the critical field $H_0=1/4$. So, while \eqref{hmc} naively suggests that the effects of the
$p_1\mapsto -p_1$ asymmetry should be maximal in the limit $H_0\ra 1/4^-$, and absent for $H_0=0$, in reality the effect vanishes in both limits and should be most observable when $H_0$ is neither too small nor too big.

\section{Concluding remarks}\news\label{sec:conc}

In this paper, we have numerically constructed Skyrmions in a simple model of frustrated ferromagnets in the regime of low applied magnetic field ($H_0<1/4$) where the
system's ground state is a spatially varying conical spiral $\svec_0(\xvec)$. These Skyrmions are spatially localized topological defects sitting on top of the ground state: as $r\ra\infty$, the spin field approaches $\svec_0(\xvec)$. They have a core width comparable to the spiral period of the ground state, rather long range tails, and possess very little symmetry. Multiple Skyrmions can bind together and form bound states with lower and lower energy per unit topological charge. In the limit $H_0\ra1/4$, unit Skyrmions connect continuously to the axially symmetric Skyrmions of the high field ($H_0\geq1/4$) regime. Finally, we analyzed the propagation of small amplitude spin waves through the conical ground state, finding strong dependence on $H_0$ and propagation direction, and, in contrast to the high field regime, total absence of a spectral gap. 

The key to deriving these results was a mathematical trick. We reinterpreted the original field theory as a {\em gauged } sigma model for $\svec$, with trivial gauge field $\Avec=0$, then changed gauge so that the conical ground state is constant, but the gauge field is non-zero. This allowed us to implement the boundary condition simply, and to make sense of the notion of topological charge. 

The results presented above are of physical interest as they suggest that Skyrmions can exist in regimes that have not so far been considered, and that such Skyrmions have rather novel properties. Skyrmions in the low field regime are actually more stable than their high field counterparts, as shown in figure \ref{Fig:varyHa}, and form bound states with distinct differences (very low symmetry, preferred orientation, spatially extended tails), which one might hope to observe in the laboratory. Finally this paper also changes the understanding of what happens in materials as we approach the critical field $H_0 = 1/4$.

Several developments of this work immediately suggest themselves. It would be interesting, and pertinent for technological applications, to study how these Skyrmions react to an applied electric field.  This question is subtle because, depending on the field's orientation relative to the ground state, it may change the ground state itself, and this effect should be included in any model of the dynamics. Another interesting dynamical question is whether the absence of a spectral gap in the dispersion relation for spin waves has a qualitative effect on the Skyrmion dynamics. One suspects this absence may amplify the effect of radiative dissipation, making Skyrmions less mobile in the low field regime. 

In this work we considered only the simplest model of frustrated ferromagnets: many other terms can be added to $E(\svec)$. One simple but interesting possibility is to add the effect of intrinsic anisotropy to the system, by including an easy axis term $-\kappa(\evec_3\cdot\svec)^2$ to the energy density. The ground state phase diagram (in the $(H_0,\kappa)$ plane) is now much more elaborate \cite{leomos}, and the task of constructing Skyrmions on top of the ground state correspondingly more challenging. Perhaps the most interesting question is whether, in the limit of large $n$, Skyrmions form regular doubly periodic lattices. If so, which field is periodic, $\svec$ or $\uvec$, or some other gauge transform of $\svec$? And what is the optimal period lattice? The precession direction of the conical ground state spontaneously breaks the rotational symmetry of the system, so there is no reason, {\em a priori} to assume, as is standard, that square or triangular lattices are energetically preferred. 

Finally, can the analysis developed here be adapted to chiral ferromagnets, where the DMI term dominates over (or replaces entirely) frustration as the stabilization mechanism?
It is striking that Schroers has also recently used gauged sigma models as a mathematical
device to analyze ferromagnets \cite{sch-magnetic}, in precisely this setting. There are some important differences between his set-up and ours. Ours is an abelian gauge theory, and our connexion (gauge field) is flat and chosen to compensate for the spatial dependence of the ground state. By contrast, Schroers's gauge theory is nonabelian (gauge group $SU(2)$) and his connexion has constant but nonzero curvature. Furthermore, the form of this connexion is dictated by the structure of the DMI term in the energy, not by the structure of the ground state. It seems likely that the mathematical device of artificially changing gauge in continuum
models of magnetic materials will prove to be of wide utility.

\subsection*{Acknowledgements}
We thank Paul Sutcliffe, Bernd Schroers and Derek Harland for useful conversations, and the anonymous referee for suggesting substantial improvements to section \ref{sec:spinwaves}.
This work was supported by the
UK Engineering and Physical Sciences Research Council through grant EP/P024688/1.

\bibliographystyle{h-physrev}
\bibliography{bibliography}

\end{document}